# A study of the new hemispherical 9-inch PMT

Fengjiao Luo[a,b,c], Yuekun Heng[a,b,c,1], Zhimin Wang[b,1], Zhonghua Qin[a,b], Anbo Yang[a,b,c], Nan Li[a,b,c], Gang Wang[a,b,d], Yan Zhang[a,b,c], Zhiyan Cai[a,b,c], Mengzhao Li[a,b,c], Haiqiong Zhang[b,c], Meihang Xu[a,b], Zhi Wu[a,b], Yuanbo Chen[a,b]

[a]State Key Laboratory of Particle Detection and Electronics, Beijing 100049, China
[b]Institute of High Energy Physics, Chinese Academy of Sciences, Beijing 100049, China
[c]University of Chinese Academy of Science, Beijing 100049, China
[d]University of South China, Hengyang 421001, China

**Abstract**

The new hemispherical photomultiplier tubes (PMTs) with 9 inch diameter from Hainan Zhanchuang Photonics Technology Co.,Ltd (HZC) have been studied. Narrow transit time spread (FWHM=2.35 ns) accompanied by small nonlinearity (750 photoelectrons at 5%) and high gain ($1\times10^7$) with good single photoelectron (PE) resolution have been observed. 11 PMTs of this type are deployed and studied in the prototype detector for JUNO at IHEP, China.

## 1. Introduction

Photomultiplier tube is widely applied in nuclear and particle physics experiments. In many large-scale scientific facilities, such as DayaBay [1], Borexino [2], Double Chooz [3] and JUNO [4], the PMT with large diameter is considered as photo-sensor detecting the weak light. The Jiangmen Underground Neutrino Observatory (JUNO) is a multipurpose neutrino experiment with a 20-thousand ton liquid scintillator detector of unprecedented 3% energy resolution (at 1 MeV) with 700 meter rock overburden located at Kaiping, Jiangmen, in southern China. It is designed to determine neutrino mass hierarchy and precisely measure the oscillation parameters by detecting reactor neutrinos from the Yangjiang and Taishan Nuclear Power Plants, and it can observe supernova neutrinos, study the atmospheric, solar neutrinos and geo-neutrinos, and perform exotic searches [4].

Following the design of the JUNO project and considering R&D requirements of each sub-system, a prototype detector was proposed to study some key technical items, including the performances of the candidate PMTs (three companies: Hamamatsu [5] from Japan, Hainan Zhanchuang Photonics Technology Co.,Ltd(HZC) [6] from China and North Night Vision Technology Co.,Ltd (NNVT) [7] from China), the liquid scintillator, electronics, waveform analysis, etc. The prototype detector for JUNO uses a stainless-steel tank of 2 m diamteter as the container. An acrylic sphere located at the tank center as the liquid scintillator (LS) vessel, and the scintillation photons from LS is viewed by 51 PMTs dipped in pure water. And 11 nine-inch new PMTs of type XP1805 are installed in the prototype detector for JUNO [8] in 2016. The PMT XP1805 is produced by Hainan Zhanchuang Photonics Technology Co., Ltd (HZC), China, with the production line imported from Photonis (France).

In this work, the key parameters of the new hemispherical PMT of XP1805 produced by HZC have been studied in detail including the quantum efficiency (QE), spectral response, single photoelectron response (SPE), dark pulse rate, linearity and time properties. The results have shown that the key parameters of the new PMTs are comparable to the 8-inch PMT R5912 [9] produced by Hamamatsu. Besides, this type of PMT can also be a new candidate option for other large-scale facilities such as LHAASO [10].

## 2. Characterization

The new PMT of type XP1805 as show in Fig.1 has a multiplier structure with a new type of high bi-alkali photocathode, and PMT XP1805 has a foil first dynode followed by seven linear focused dynode. In this section, results of critical parameters of the PMTs are described, including the quantum efficiency(QE), single photoelectron (SPE) response, gain, anode dark pulse rate, linearity of anode signal and the Transit Time Spread (TTS); We will show some major characterizations of 16 samples of the kind PMT at the same time.

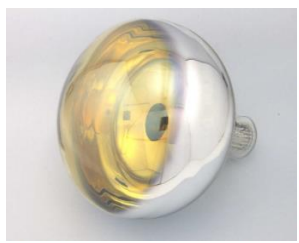

Fig.1 New hemispherical 9-inch PMT XP1805.

*2.1. Quantum efficiency*

To maximize the detector light yield (number of photoelectrons per unit energy deposited) and reach the unprecedented 3% energy resolution of JUNO, high quantum efficiency (QE) is necessary. The schema of the test for QE measurement is shown in Fig.2. QE is measured in DC photocurrent mode. Light from the deuterium lamp is guided through a monochromator to PMT. The focusing-dynode structure is shortcut as a combined stage and the typical recommended photocathode-1st dynode voltage increment is applied between the combined stage and photocathode. Fig.3 (a) shows the curves of the cathode current as a function of the operating voltage of 3 samples of PMT XP1805 when wavelength is at 410 nm. Photocathode is connected to a pico-ampere meter before ground. A calibrated photodiode (PD) from Hamamatsu is used as a reference photo detector to provide an absolute calibration of PMT. The QE of PMT can be expressed as:

$$\eta_{PMT} = \eta_{PD} \frac{\eta_{PMT}}{\eta_{PD}} \qquad (1)$$

$$\eta_{PMT} = \eta_{PD} \frac{I_{PMT}}{I_{PD}} \qquad (2)$$

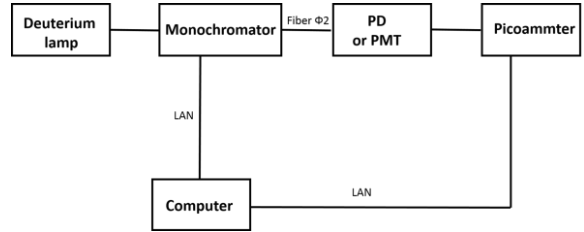

Fig. 2. The schema of experimental setup for QE measurement.

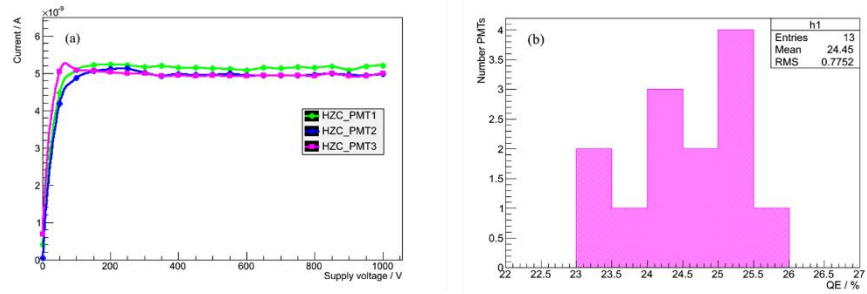

Fig. 3. (a) Cathode current as a function of the collecting voltage; (b) QE histogram for 13 PMTs.

From Fig.3 (a), the cathode output current as a function of the operating voltage of 3 PMTs will reach plateau at the voltage around 200 V, which means the photoelectrons have been collected totally. The dark current has been subtracted out in Fig.3 (b); Keep intensity of LED the same, do the same measurement with PD. According to equation (1) and (2), QE of PMTs will be obtained.

With the same setup, we tested the remaining new PMTs for JUNO prototype. And the results are shown in Fig.3 (b), the values are almost more than 23%, average QE is around 24.5%.

On the other hand, for JUNO, the neutrino is measured by the inverse-beta decay, detecting photons from the scintillator by the PMTs and it is crucial to match the spectrum between the emission spectrum of liquid scintillator and response spectrum of PMTs. In order to study the spectral response of the PMTs, scan wavelength by the monochromator; and then measure the QE of PD or PMT under different wavelength. We take 2 PMTs for example, the spectral responses are shown in Fig.4.

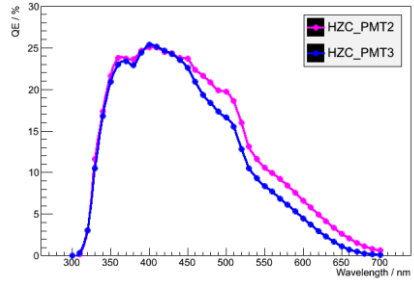

Fig. 4. The spectral response of PMTs.

The spectral response is related to photocathode and window materials of PMT. From Fig.4, the two spectral responses are consistent; the maximum of QE is all around 400nm of the PMTs which can be matched well with the emission wavelength of liquid scintillator, so the photons from liquid scintillator can be detected by PMTs efficiently; there is almost no response of PMTs when the wavelength is longer than 700 nm. The study of spectral response of PMTs can help manufacturers to understand the photocathode and window materials of PMTs [11] and on the other hand, the LED at a specific wavelength which the PMTs are insensitive to can be adopted as auxiliary facilities in photodetector project without causing spurious signal, such as infrared LED with CCD as a monitor in JUNO [4].

*2.2. SPE response*

The PMTs should be high gain to distinguish noise from signal because of the slight neutrino signal [4]. It is necessary to measure the gain of PMTs as an evaluated parameter for JUNO PMTs. The optimized schema of the high voltage divider to study the anode performances of PMT is shown in Fig.5 (a).

With the optimized high voltage divider, the overshoot of waveform is lower than 1% [12]. Fig.5 (b) shows the waveform of SPE.

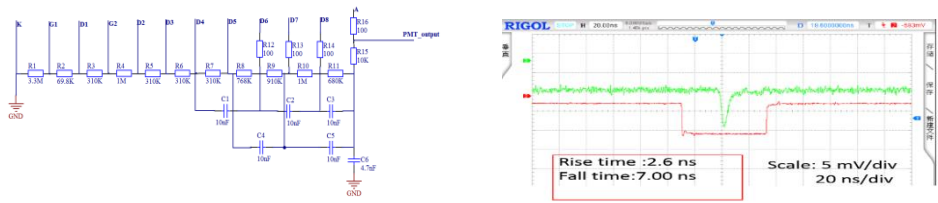

Fig. 5. (a) Optimized high voltage divider; (b) Waveform of SPE: the scale with 5mV/div and 20 ns/div.

The gain, P/V (Peak to Valley), energy resolution can be obtained from Single Photoelectron (SPE). The schema of the measurement system of SPE is shown in Fig.6 (a). PMT and LED are put in the dark box, where LED is driven by pulse generator. At the same time, a synchronized gate from pulse generator is sent to a low threshold discrimination (LTD) as a trigger for QDC V965. The voltage of PMT is provided by a CAEN Mod.A1733P in a CAEN SY4527 Crate. Anode signals are measured by QDC V965. Fig.6 (b) shows the SPE of one PMT [13]; the P/V ratio is 2.53 and the energy resolution is around 24%.

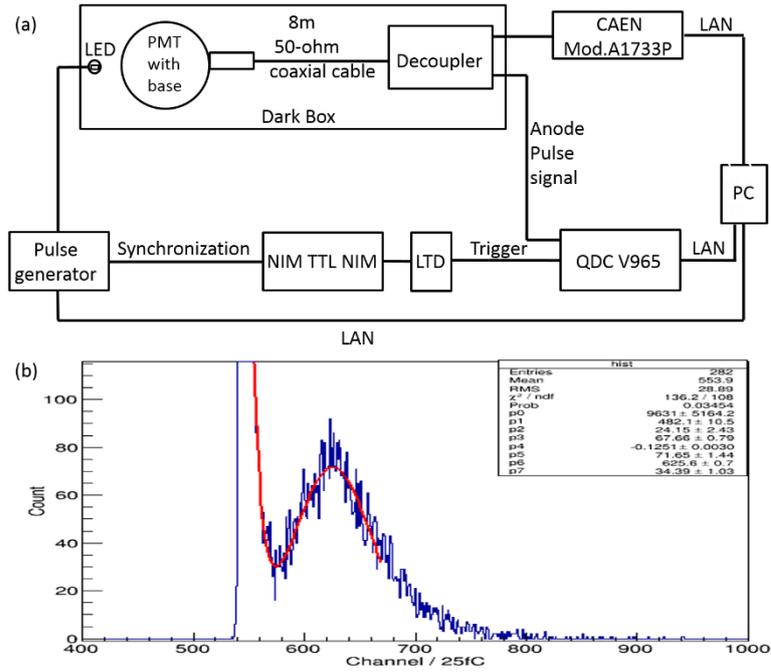

Fig. 6 (a) The schema measurement system of SPE; (b) The SPE of one PMT-XP1805.

The curve of HV verse gain is shown in Fig.7 (a) in a semi-log scale. When the voltage is above 2150 V, the gain is higher than $1\times10^7$. The curve is consistent with the exponential distribution from the range of 2050 V and 2700 V. The P/V ratio is more than 2, as shown in Fig.7 (b).

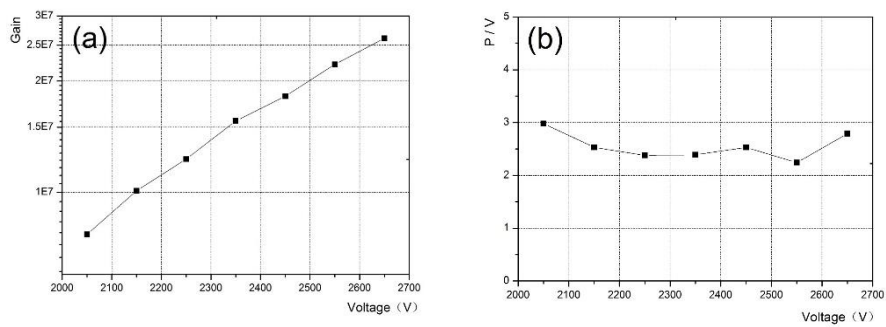

Fig. 7. (a) The HV verse gain curve; (b) The HV verse P/V ratio.

The distributions of voltage and P/V ratio at the gain of $1\times10^7$ of 16 PMTs are shown in Fig.8 (a) and Fig.8 (b). From Fig.8, we can know that the new PMTs can reach the gain at $1\times10^7$, and the voltages are lower than 2800 V except two PMTs. The P/V of PMTs with the gain at $1\times10^7$ is between 1.2 and 3. The performances of the new PMT in SPE response are almost consistent. For the project, considering the aging of PMTs, the gain of PMT will have a variation [14]. To keep the same gain at $1\times10^7$, the voltage will be increased as the run of the project. So 2 PMTs with high voltage around 2800 V are not installed in JUNO prototype.

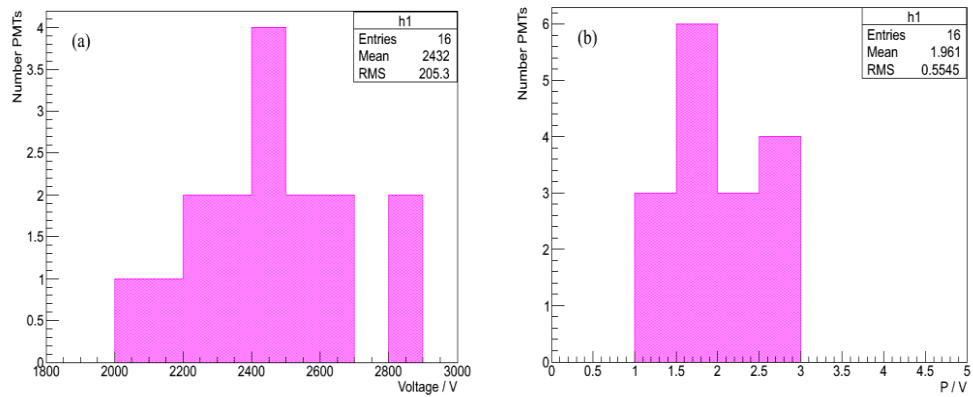

Fig. 8. (a) The distributions of voltage when the gain is around $1 \times 10^7$ of 16 PMTs;
(b) The distributions of P/V ratio when the gain is around $1 \times 10^7$ of 16 PMTs.

*2.3. Anode dark count rate*

Low dark count rate with high gain of PMT is one of the key requirement for nuclear and particle physics experiment; Dark count rate increases with an increasing supplied voltage. Dark count rate may be categorized by cause as follow: thermionic emission, leakage current and field emission [15].

The setup for the dark count rate is shown in Fig.9. The experiment is performed at room temperature and kept gain at $1 \times 10^7$. The PMT is placed in the dark box before operation to reduce any effects the ambient light may have. The anode dark count rate of one PMT is shown in Fig.10. Anode dark count rate at different thresholds (2 mV, 3 mV, 4 mV, 5 mV) is shown in Fig10 (a).

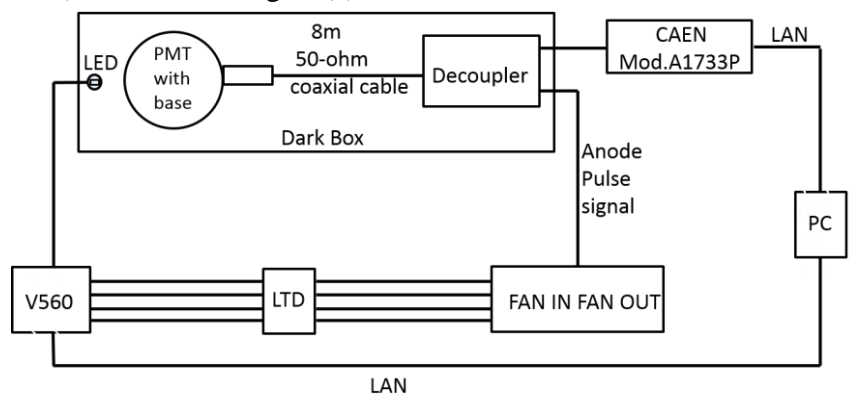

Fig. 9. The schema of pulse dark rate.

The anode dark count rate decreases significantly as time goes on. After several hours, the dark rate reaches stable and keeps at 8 kHz with 3 mV threshold. Besides, with the same setup, we have also tested all of the 16 PMTs and the results are shown in Fig.10 (b). There is one PMT with abnormal dark rate around 290 kHz and one PMT reached 32 kHz, others are less than 10 kHz. The PMTs with anode dark count rate less than 10 kHz can satisfy JUNO project.

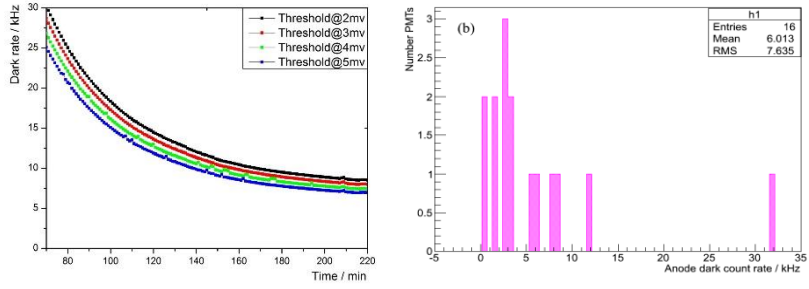

Fig. 10. (a) Anode dark count rate for PMT-XP1805-00151 with different thresholds; (b) Results of anode dark count rate for 9-inch PMTs.

## 2.4. Linearity

In most neutrino experiments, any one PMT will not detect more than thousands photons. When the detected photons up to a few hundred are detected, the output of PMT is proportional to the number of photons detected. However, some of the signal events such as cosmic ray in JUNO would be expected to deposit much energy within tens of meters of individual PMT, and then the PMT response will not be proportional. For the precise charge measurement, the charge non-linearity are required to be studied.

The measurement schema is shown in Fig.11. A pulse generator can drive two blue LEDs in succession and then make them emitting together. The light intensity of each LED can also be regulated separately to cover the whole PMT dynamic range. The light pulse from LED1 signed as A, the light pulse (B) from LED2 and a sum of light pulse C from LED1 and LED2 flashing simultaneously. At the same time, a FADC system with 1GHz sampling will record the waveforms respectively. If the PMT is ideally linear, we can obtain,

$$C = A + B \tag{2}$$

However, as known, the PMT response has a nonlinear effect. Than the deviation of the linearity can be expressed as,

$$Linearity = \frac{C - (A_{corrected} + B_{corrected})}{A_{corrected} + B_{corrected}} \tag{3}$$

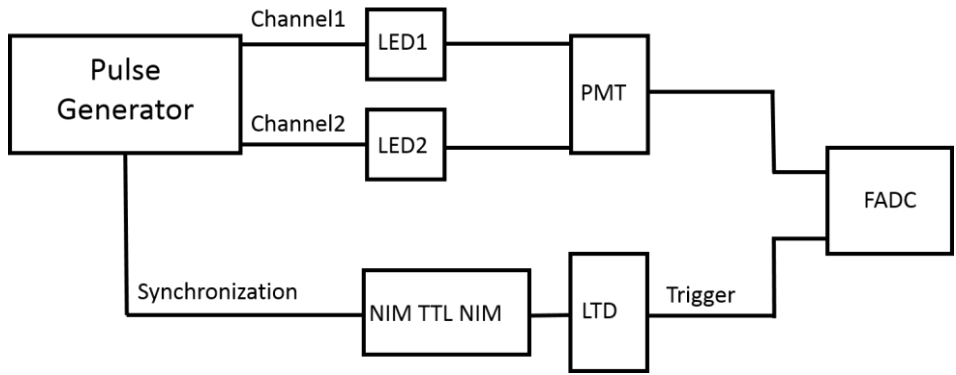

Fig. 11. The schema of measurement system of linearity.

Where $A_{corrected}$ ($B_{corrected}$) is the corrected A (B) according to the measured nonlinearity effect in the lower intensity range. The measurement results are demonstrated in Fig.12. The figure shows that the number of photoelectrons of PMT can reach 750 as the nonlinearity of 5%.

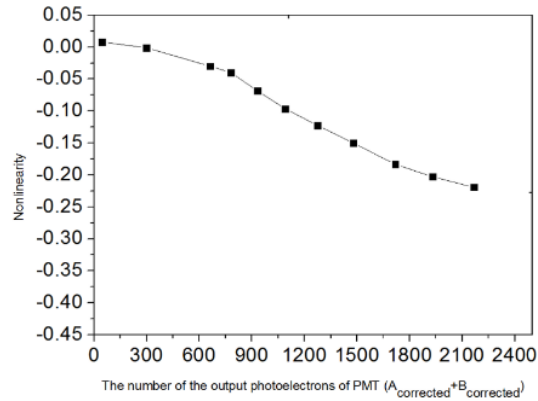

Fig.12. Nonlinearity of the PMT.

*2.4. Transit time spread*

Transit time spread (TTS) is defined as full width at half maximum (FWHM) of a signal's transit time distribution which is very important for vertex and track reconstruction of the neutrino. Photons arriving at different positions on the photocathode will result in photoelectrons with different transit time, which will contribute to the final TTS of signal. Normally in the PMT manufactures, optimized electro-optical simulation and structure design should have been executed to suppress such effect. The TTS setup is almost the same as Fig.6 (a), the difference is that waveforms are sampled by oscilloscope RIGOL DS6104 with 5Gs/s instead of QDC.

Fig.13 (a) shows the transit time histogram of SPE at the gain of $1 \times 10^7$ measured with point-like light. TTS of SPE at the gain of $1 \times 10^7$ is around 2.323ns. Besides, we study the TTS vs high voltage as shown in Fig.13 (b). TTS improves in inverse proportion to the square root of the supply voltage under 2450 V. When the supply voltage reached 2500 V, TTS increased instead which is caused by the noise of PMT.

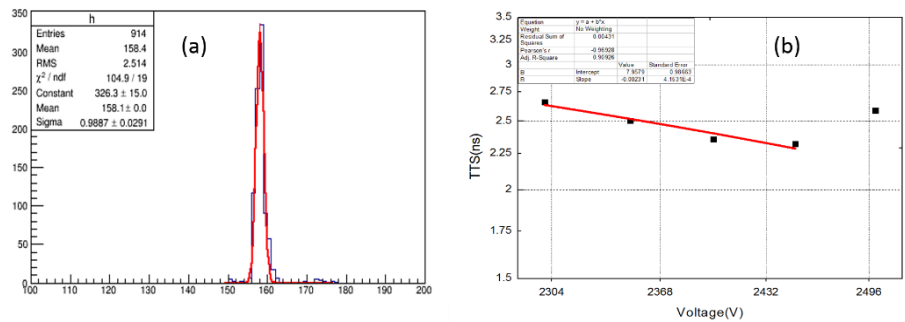

Fig.13. (a) Transit time histogram of SPE at the gain of $1 \times 10^7$, and transit time spread is signified by FWHM which is equal to $2.35 \times$ sigma; (b) Transit time spread versus voltage with log-log coordinate.

## 3. Summary

The detailed characterizations of the new 9 inch PMT of XP1805, such as the cathode properties, the anode signal response, and the time properties are studied in this paper. In general, the performances of the new PMT are good. Its QE is almost around 25% at the wavelength of 410 nm and the spectral response are matched well with emission wavelength of the liquid scintillator; Its gain of $1 \times 10^7$ can be reached with the voltage less than 2800 V; It has lower anode dark count rate and fast time properties of rise time 2.0 ns, fall time 7.0 ns and TTS 2.3 ns. We tested 11 PMTs of XP1805 in the JUNO prototype and they can run stably with the above properties. The general features of this new PMT can be of interest for the applications in the neutrino physics and astrophysics experiments, which has the demanding to detect photons in large area.

## Acknowledgement

The work was supported by the Strategic Priority Research Program of the Chinese Academy of Sciences (Grant No.XDA10011200) and High Energy Physics Experiment and Detector R&D (By Chinese Academy of Sciences).